\documentclass[copyright,creativecommons]{eptcs}
\providecommand{\event}{MSFP 2022} 
\usepackage{breakurl}             
\usepackage{underscore}           

\usepackage{amsmath,amssymb,amsthm}
\usepackage{xcolor}
\usepackage{relsize}

\usepackage{bm}
\usepackage{comment}

\newcommand{\norm}[1]{\left\lVert#1\right\rVert}
\newcommand{\deriv}[1]{{#1}'}

\newcommand{\aff}[1]{{#1}^{[1]}}

\newcommand{\dx}{\mathit{dx}}
\newcommand{\dy}{\mathit{dy}}
\newcommand{\fx}{\mathit{fx}}
\newcommand{\fpx}{\mathit{f'\!x}}
\newcommand{\fpxa}{\mathit{f'\!xa}}
\newcommand{\gfx}{\mathit{gfx}}
\newcommand{\gpfx}{\mathit{g'\!fx}}
\newcommand{\gpfxa}{\mathit{g'\!fxa}}

\newcommand{\comp}{\circ}
\newcommand{\lcomp}{\bullet}

\newcommand{\pderiv}[2]{\frac{\partial{#1}}{\partial{#2}}}

\newcommand{\const}[1]{K_{#1}}
\newcommand{\lift}[1]{\dot{#1}}
\renewcommand{\lift}[1]{\hat{#1}}

\newcommand{\blank}{\textrm{\textvisiblespace}}
\newcommand{\Let}{\textbf{let }}
\newcommand{\ULet}{\textbf{\underline{let} }}
\newcommand{\In}{\textbf{ in }}

\newcommand{\dup}{\mathrm{dup}}
\newcommand{\rep}{\mathrm{rep}}
\newcommand{\zipapp}{\Pi}
\newcommand{\unzip}{\mathit{unzip}}
\newcommand{\zip}{\mathit{zip}}
\newcommand{\zipWith}{\mathit{zipWith}}

\newcommand{\fst}{\pi_1}
\newcommand{\snd}{\pi_2}

\DeclareMathOperator{\map}{\mathrm{map}}

\newcommand{\id}{\mathrm{id}}

\newcommand{\distrib}{\mathit{distrib}}
\newcommand{\idistrib}{\distrib^{-1}}

\newcommand{\mult}{\cdot}
\newcommand{\mmult}{\star}
\newcommand{\tensor}{\ast}

\DeclareMathOperator{\red}{\mathrm{red}}

\newcommand{\Term}[1]{\mathrm{Term}(#1)}
\newcommand{\Hom}[2]{L({#1}, {#2})}
\renewcommand{\Hom}[2]{[{#1} \Rightarrow {#2}]}
\renewcommand{\Hom}[2]{{#1} \rightarrow_1 {#2}}
\renewcommand{\Hom}[2]{{#1} \multimap {#2}}
\newcommand{\Aff}[2]{{#1} \rightarrow_{\leq 1} {#2}}
\newcommand{\Iso}[2]{{#1} \longleftrightarrow_1 {#2}}

\newcommand{\Bilin}[3]{{#1} \times {#2} \rightarrow_2 {#3}}
\newcommand{\Func}[2]{{#1} \rightarrow {#2}}

\newcommand{\biprod}[2]{\oplus_{#1} {#2}}
\renewcommand{\prod}[2]{{#2}^{#1}}
\renewcommand{\coprod}[2]{{#2}^{(#1)}}
\renewcommand{\coprod}{\prod}
\newcommand{\scan}{\mathrm{scan}}

\newcommand{\seg}[1]{\ulcorner #1 \urcorner}
\renewcommand{\seg}[1]{\mathbf{#1}}

\newcommand{\inj}[2]{\iota_{#1}^{#2}}
\newcommand{\proj}[2]{\pi_{#1}^{#2}}

\newcommand{\inverse}[1]{#1^{-1}}
\newcommand{\grad}{\nabla}

\newcommand{\power}[1]{\mathrm{power}^{#1}}
\renewcommand{\power}[2]{{#2}^{#1}}

\newcommand{\Nat}{\mathbb{N}}
\newcommand{\Int}{\mathbb{Z}}
\newcommand{\Real}{\mathbb{R}}
\newcommand{\Complex}{\mathbb{C}}

\newcommand{\eval}[1]{\mathrm{eval}^{[1]}[\![{#1}]\!]}
\newcommand{\evalz}[1]{\mathrm{eval}^{(0)}[\![{#1}]\!]}
\newcommand{\evalf}[1]{\mathrm{eval}^{(\mathrm{f})}[\![{#1}]\!]}

\newcommand{\evalreverse}[1]{{#1}^{[1r]}}

\newcommand{\bi}[1]{\bar{#1}}

\newcommand{\dual}[1]{{#1}^*}
\newcommand{\dualvec}{\mathit{dual}}
\newcommand{\adjoint}[1]{{#1}^\dagger}
\newcommand{\adj}[1]{{#1}^*}

\newcommand{\bra}[1]{\mbox{\raisebox{3pt}{${}_{\pmb{\langle}}\mkern 1mu$}}{#1}}      
\newcommand{\ket}[1]{{#1}\mbox{\raisebox{3pt}{$\mkern 2mu{}_{\pmb{\rangle}}$}}}      
\newcommand{\ibra}[1]{| #1 |}
\newcommand{\iket}[1]{| #1 |}
\newcommand{\transpose}[1]{{#1}^T}
\newcommand{\assoc}{\mathit{assoc}}
\newcommand{\iassoc}{\assoc^{-1}}
\includecomment{note}
\newlength{\dhatheight}

\newcommand{\todo}[1]{\textcolor{red}{#1}}
\renewcommand{\todo}[1]{}
\newcommand{\dv}{\mathit{dv}}

\newtheorem{theorem}{Theorem}[section]

\newtheorem{proposition}[theorem]{Proposition}
\theoremstyle{definition}
\newtheorem{example}[theorem]{Example}
\newtheorem{definition}[theorem]{Definition}

\title{Combinatory Adjoints and Differentiation}

\author{Martin Elsman \institute{DIKU, U.~Copenhagen} \email{mael@diku.dk}
\and Fritz Henglein \institute{DIKU, U.~Copenhagen} \email{henglein@diku.dk}
\and Robin Kaarsgaard \institute{U.~Edinburgh} \email{Robin.Kaarsgaard@ed.ac.uk}
\and Mikkel Kragh Mathiesen \institute{DIKU, U.~Copenhagen} \email{mkm@di.ku.dk}
\and Robert Schenck \institute{DIKU, U.~Copenhagen} \email{rschenck@di.ku.dk}
}

\def\titlerunning{Combinatory Adjoints and Differentiation}
\def\authorrunning{Elsman, Henglein, Kaarsgaard, Mathiesen, Schenck}

\newcommand{\memo}[1]{{\color{red}MEMO: #1}}
\renewcommand{\memo}[1]{}

\begin{document}

\maketitle

\begin{abstract}


We develop a compositional approach for automatic and symbolic differentiation based on categorical constructions in functional analysis where derivatives are linear functions on abstract vectors rather than being limited to scalars, vectors, matrices or tensors represented as multi-dimensional arrays.

We show that both symbolic and automatic differentiation can be performed using a differential calculus for generating linear functions representing Fr\'echet derivatives based on rules for primitive, constant, linear and bilinear functions as well as their sequential and parallel composition.  Linear functions
are represented in a combinatory domain-specific language.  

Finally, we provide
a calculus for symbolically computing the adjoint of a derivative without using matrices, which are too inefficient to use on high-dimensional spaces. The resulting symbolic representation of a derivative retains the data-parallel operations from the input program.  The combination of combinatory differentiation and computing formal adjoints turns out to be behaviorally equivalent to reverse-mode automatic differentiation. In particular, it provides opportunities for optimizations where matrices are too inefficient to represent linear functions.

\end{abstract}

\section{Introduction}

Automatic differentiation (AD) \cite{griewank2008evaluating} is the
discipline of computing derivatives for functions given by programs.
It is used in gradient-based optimization, neural networks,
probabilistic inference \cite[sec. 4]{adsurvey} and has numerous
applications in computer vision, natural language processing,
computational science, bioinformatics, quantitative finance,
computational economics, and in many other areas.  For example,
backpropagation, which is used in machine learning to train neural
networks, is an instance of reverse mode AD.  Building tools to
implement and compute derivatives from programs automatically,
efficiently, and precisely, has far-reaching impact potential.

\todo{
higher-order, generality, performance characteristics (compared to
manual differentiation), parallelism preservation, higher-degrees,
balancing forward/backward-mode (hybrid), utilising algebraic
properties (often derivatives code is simpler than the source
code). Data-agnostic...}

\subsection{Contributions}

\todo{Notice: Once we are clear about the innovation part, we
  can properly put each of the related work items in perspective.}
In this paper we develop a general framework for expressing and reasoning about functions, their derivatives and the adjoints of these in combinatory form.

We make the following novel contributions:
\begin{itemize}
\item We present a general framework for constructing Hilbert spaces.
The constructions freely combine tensor products and direct sums.
Direct sums generalize both homogeneous data types, such as order-$k$ tensors (scalars, vectors, matrices, and so on) and
inhomogeneous types such as tuple and record types.  Abstract tensor products express tensor decomposition of matrices, which 
are asymptotically more efficient than using matrices for low-rank matrices.
\item We identify five general differentiation rules for calculating Fr\'echet derivatives, which represent derivatives as linear functions, by structural recursion on functions given in combinatory form.  The combinatory form of a function thus distills its differential properties. Intuitively, differentiating a function in point-full notation consists mostly of (implicitly) turning it into combinatory form.
\item We exhibit the generalized product rule, which is applicable to arbitrary bilinear functions operating on spaces of any dimension as a general rule not previously exploited.  Bilinear functions on high-dimensional data are common, including matrix multiplication, outer product, dot product, zip (Hadamard) product and any composition of a linear function with a bilinear function.  To differentiate a bilinear function we only need to know that it is bilinear since its derivative is expressed in terms of itself.
\item We provide affine interpretation of a function in combinatory form, which computes both the output value of a function at a given input and returns a symbolic (term) representation of its Fr\'echet derivative.  Symbolic rather than functional representations facilitate optimization using (multi)linear and tensor algebra equalities.
\item We further demystify reverse-mode automatic differentiation by identifying its essence as symbolically computing the adjoint of the Fr\'echet derivative in combinatory form.  The adjoint of a linear function $f$ is a representation of its transpose, the continuation passing style version of $f$.  
In adjoints linear continuations are represented by their duals, ordinary first-order vectors, which facilitates and explains how a linear function can be executed efficiently in reverse.
\item We provide an adjoint calculus for symbolically calculating the adjoints of linear functions in combinatory form.  We identify \emph{relational reduction} and  \emph{tensor contraction} as natural parallel linear operations since they provide their own adjoints.
\item We illustrate how combinatory differentiation and combinatory adjoint calculation can be used to derive the backpropagation code for neural networks such that all data parallelism is preserved.
\end{itemize}
More speculatively, we believe our combinatory setting is useful for a differential and adjoint calculus on functions and linear functions.  The Hilbert space setting seems to provide a promising setting in which both database and analytic functions can be specified, differentiated and reversed by taking adjoints.

\subsection{Outline}

We assume basic familiarity with functional analysis, which, as a framework, generalizes both multivariate and tensor calculus by operating on arbitrary elements of structured vector spaces instead of restricting them to tuples of scalars or multi-dimensional arrays that represent tensors.  The relevant notions are introduced in the remainder of this and the next section.

In Section~\ref{combinatoryform.sec} we informally present a list of primitive, constant, linear and bilinear analytic functions that can be combined freely by sequential and parallel composition to complex analytic functions in point-free notation.  In Section~\ref{differential-calculus} we then formulate a calculus for symbolically differentiating 
functions in combinatory form such that the derivatives of parallel functions are rendered in point-free notation, as parallel linear functions.
In Section~\ref{affine-interpretation.sec} we show how this gives rise to affine interpretation of an analytic function in combinatory form: The interpreter
returns not only the value of a function on its input, but also a compact combinatory representation of its derivative whose size is largely independent of the dimensionality of the vector spaces involved.
In Section~\ref{adjoints} we show how the inner product operator can be used to uniquely represent linear continuations by ordinary first-order vectors.  This gives rise to symbolically computed adjoints, which run a linear function efficiently ``in reverse'' and thus provide reverse-mode AD.  We illustrate combinatory differentiation on neural networks in Section~\ref{applications.sec} and discuss related work in Section~\ref{discussion}.

\subsection{Background}
\label{background.sec}

\todo{
Derivatives (Leibniz, Jacobi, Fr\'echet) \\
Polynomial ring, ideal, dual numbers \\
Ring, filed, module, vector space, Banach space, inner product space, Hilbert space, Euclidean space, unitary space \\
Linear map, semilinear function, adjoint \\
Category, functor
}

\begin{definition}[Fr\'echet derivative]
For a function $f: V \rightarrow W$ on Banach spaces $V, W$, the \emph{linear function} $A \in \Hom{V}{W}$ is the \emph{Fr\'{e}chet derivative of $f$ at $v$} if it satisfies
$$f(v + \dv) \approx f(v) + A(\dv),$$ that is
\begin{equation*}
\lim_{||\dv||_V \rightarrow 0} \frac{||f(v + \dv) - (f(v) + A(\dv))||_W}{||\dv||_V} = 0 
\end{equation*}
where $|| \ldots ||_U$ is the norm that comes with the Banach space $U$.

The \emph{Fr\'{e}chet derivative of $f : \Func{V}{W}$} is the partial function $\deriv{f}: \Func{V}{(\Hom{V}{W})}$ that maps a vector $v \in V$ to the Fr\'{e}chet derivative of $f$ at $v$.
\end{definition}
See Appendix~\ref{derivatives.sec} for other notions of derivatives, including Gateaux derivatives. 

%

\section{Sets and spaces}
\label{setsandspaces.sec}

We provide general constructions for defining \emph{inner product spaces} over $\Real$ and their implicit completions to real Hilbert spaces.  These provide a model  of symbolic derivatives as Fr\'echet derivatives.

An inner product space over $\Real$ is a vector space $V$ over $\Real$ equipped with an inner product $$\odot: \Bilin{V}{V}{\Real}$$ that is symmetric, $v_1 \odot v_2 = v_2 \odot v_1$, and positive definite, $v \odot v > 0$ for all $v \neq 0$.
A real Hilbert space is an inner product space over $\Real$ that is also a complete metric space with respect to the distance function $d(v, w) = \norm{v - w}$ where $\norm{v} = \sqrt{v \odot v}$.

A continuous function $f: \Func{V}{W}$ on real Hilbert spaces $V, W$ is \emph{linear} if $f(u + v) = f(u) + f(v)$ and $f(k \cdot v) = k \cdot f(v)$; we
write $f: \Hom{V}{W}$ if $f$ is continuous and linear.

A continuous binary function $\diamond: \Func{U \times V}{W}$ is \emph{bilinear} if $(u \diamond): \Hom{V}{W}$ and $(\diamond v): \Hom{U}{W}$ are linear for all $u \in U, v \in V$ where $(u \diamond)$ and $(\diamond v)$ are defined by $(u \diamond)(v) = u \diamond v = (\diamond v)(u)$.  We write $f: \Bilin{U}{V}{W}$ if $f$ is continuous and bilinear.

Proviso: Henceforth all functions will implicitly be continuous.


\subsection{Sets}

We provide a language for defining \emph{index sets}.  These are used to construct direct sum spaces.
\begin{eqnarray*}
X, Y & ::= & \seg{n} \mid X \times Y \mid X + Y
\end{eqnarray*}
where $n \in \Nat$, $\seg{n}$ is the initial segment $\{ 1, \ldots, n \}$ of natural numbers; $S \times T$ and $S + T$ the Cartesian product, respectively disjoint union of $S$ and $T$.  

The constructible index sets are finite, which ensure that the constructions are metrically complete.  We believe the theory, being essentially algebraic,
can be extended to infinite denumerable sets.  We stick to finite index  sets and thus finite-dimensional Hilbert spaces in this paper, however.



\subsection{Spaces}

Below we provide constructions for Hilbert spaces generated by the following terms:
\begin{eqnarray*}
U, V, W & ::= & 0 \mid K \mid \biprod{x \in X}{V_x} \mid V \otimes W
\end{eqnarray*}
where $V_x$ may depend on $x \in X$. 

\subsubsection{Atomic spaces}

The trivial vector space $0$ consists of the single element $0$.

$K$ stands for the underlying field of our vector spaces, here $\Real$. Its elements are the elements of $\Real$ as a field. 
Its operations as a vector space are the corresponding field operations.

\subsubsection{Direct sum space}

The Hilbert space $V = \biprod{x \in X}{V_x}$ for denumerable $X$ is the
\emph{(external) direct sum} of a family of Hilbert spaces $V_x$ indexed by $x \in X$.  Its elements are
maps $m$ from $X$ such that $m(x) \in V_x$ and $\sum_{x \in X} (m(x) \odot_{V_x} m(x)) < \infty$. We write $m_x$ for the result of applying the map to highlight that $x$ is an element of an index set, not a vector. Its operations are defined by component-wise lifting, where the inner product is
\begin{eqnarray*}
(v \odot_V v') & = & \sum_{x \in X} (v_x \odot_{V_x} v'_x)
\end{eqnarray*}
The summation is defined since $\sum_{x \in X} (v_x \odot_{V_x} v'_x) \leq \sum_{x \in X} (v_x \odot_{V_x} v_x) + \sum_{x \in X} (v'_x \odot_{V_x} v'_x) < \infty$.  
Note it is trivially well-defined for finite $X$.

$V$ comes with linear injection and projection functions
\begin{eqnarray*}
\inj{y}{X} & : & \Hom{V_y}{\biprod{x \in X}{V_x}} \\
\proj{y}{X} & : & \Hom{\biprod{x \in X}{V_x}}{V_y}
\end{eqnarray*}
for $y \in X$, and the \emph{zipped apply} operator
\begin{eqnarray*}
\zipapp_{x \in X} f_x & : & \Func{\biprod{x \in X}{V_x}}{\biprod{x \in X}{W_x}}
\end{eqnarray*}
for a family of functions $f_x \in \Func{V_x}{W_x}$ indexed by $x \in X$.
They satisfy
\begin{eqnarray*}
\proj{y}{X} \comp \zipapp_{x \in X} f_x \comp \inj{y}{X} & = & f_y \\
\proj{z}{X} \comp \zipapp_{x \in X} f_x \comp \inj{y}{X} & = & 0_{yz} \qquad\mbox{if } y \neq z
\end{eqnarray*}
where $0_{yz} : \Hom{V_y}{W_z}$ maps all $v_x \in V_x$ to $0 \in W_z$.
The zipped apply operator preserves linearity, that is
\begin{eqnarray*}
\zipapp_{x \in X} f_x & : & \Hom{\biprod{x \in X}{V_x}}{\biprod{x \in X}{W_x}}
\end{eqnarray*}
for $f_x : \Hom{V_x}{W_x}$.  A special case of this is
\begin{eqnarray*}
\Delta & : & \Hom{\biprod{x \in X}{(\Hom{V_x}{W_x})}}{(\Hom{\biprod{x \in X} V_x}{\biprod{x \in X} W_x})}
\end{eqnarray*}
defined by
\begin{eqnarray*}
\Delta(\oplus_{x \in X} f_x)(\oplus_{x \in X} v_x) & = & \oplus_{x \in X} (f_x(v_x))
\end{eqnarray*}
which will later play the role of gathering derivatives acting on the individual differentials of a collection into a derivative that acts on all differentials in parallel.

We write $V_1 \times \ldots \times V_n$ or $V_1 \oplus \ldots \oplus V_n$ for
$\biprod{i \in \seg{n}}{V_i}$.  In particular,
$V_1 \times V_2 = V_1 \oplus V_2 = \biprod{i \in \seg{2}}{V_i}$ is the direct sum of $V_1$ and $V_2$, whose elements are the pairs $(v_1, v_2)$ such that $v_1 \in V_1$ and $v_2 \in V_2$.

\subsubsection{Copower space}

For set $X$ and space $V$, the \emph{copower}
$\coprod{X}{V}$ is the direct sum, where each space in the family is the same $V_x = V$:
$$\coprod{X}{V} = \biprod{x \in X}{V}.$$
As special cases we have $\prod{\seg{n}}{\Real}$ as the space of $n$-ary vectors of scalars.  In particular, $V \times V = \prod{\seg{2}}{V}$.
Note that the exponents are sets, not numbers.  This is reflected in the notation $\prod{\seg{n}}{\Real}$: an element is a finite map $m$ from $\seg{n}$ to $\Real$, which can conveniently be written using tuple notation $(m_1, \ldots, m_n)$. For example $(5, 8, 22) \in \prod{\seg{3}}{\Real}$ is syntactic sugar for $\{ 1 \mapsto 5, 2 \mapsto 8, 3 \mapsto 22 \}$.

For relation $R \subseteq X \times Y$ where $X, Y$ are finite we define \emph{relational reduction}
\begin{eqnarray*}
\red_R & : & \Hom{\coprod{X}{V}}{\coprod{Y}{V}} \\
(\red_R (v))_y & = & \sum_{(x, y) \in R} v_x
\end{eqnarray*}


Many useful functions can be defined in terms of relational reduction.
Let $Y \subseteq X$ be finite.  The functions
\[ \begin{array}{rcll}
\prod{X}{f} & : &  \Func{\coprod{X}{V}}{\coprod{X}{W}} & \mbox{if } f : \Func{V}{W} \\
\rep_Y & : & \Hom{V}{\coprod{X}{V}} \\
\sum_Y & : & \Hom{\coprod{X}{V}}{V} \\
+ & : & \Hom{\prod{\seg{2}}{V}}{V} \\
\dup & : & \Hom{V}{\prod{\seg{2}}{V}} \\
\scan_n & : & \Hom{\prod{\seg{n}}{V}}{\prod{\seg{n}}{V}} \\
\langle f_y \rangle_{y \in Y} & : & \Hom{U}{\biprod{y \in Y}{V_y}} & \mbox{if } f_y : \Hom{U}{V_y} \\
{}[ g_x ]_{x \in X} & : & \Hom{\biprod{x \in X}{V_x}}{W} & \mbox{if } g_x : \Hom{V_x}{W}
\end{array} \]
are defined by
\begin{eqnarray*}
\prod{X}{f} & = & \zipapp_{x \in X} f \\
\rep_Y & = & \red_{\seg{1} \times Y} \lcomp \inj{1}{\seg{1}} \\
\sum_Y & = & \proj{1}{\seg{1}} \lcomp \red_{Y \times \seg{1}} \\
+ & = & \sum_{\seg{2}} \\
\dup & = & \rep_{\seg{2}} \\
\scan_n & = & \red_{\{(i, j) \mid 1 \leq i \leq j \leq n \}} \\
\langle f_y \rangle_{y \in Y} & = & \zipapp_{y \in Y} f_y \lcomp \rep_Y \\
{}[ g_x ]_{x \in X} & = & \sum_X \lcomp \zipapp_{x \in X} g_x
\end{eqnarray*}

\subsubsection{Tensor product space}

$W = U \otimes V$ is the tensor product space of $U$ and $V$. Its finite elements are the formal terms generated by
\begin{eqnarray*}
w & ::= & 0 \mid k \cdot w \mid w_1 + w_2 \mid u \otimes v
\end{eqnarray*}
where $k \in \Real, u \in U, v \in V$ that are identified modulo the vector space axioms and the equalities
\begin{eqnarray*}
(k \cdot v) \otimes w = k \cdot (v \otimes w) = v \otimes (k \cdot w) \\
(v_1 + v_2) \otimes w = (v_1 \otimes w) + (v_2 \otimes w) \\
v \otimes (w_1 + w_2) = (v \otimes w_1) + (v \otimes w_3).
\end{eqnarray*}
We write $[w]_\otimes$ for the equivalence class of $w$ under these equalities and define
\begin{eqnarray*}
0_{W} & = & [0]_\otimes \\
v_1 +_{W} v_2 & = & [v_1 + v_2]_\otimes \\
k \cdot_{W} v & = & [k \cdot v]_\otimes
\end{eqnarray*}

$W$ is metrically complete for finite-dimensional $U, V$; otherwise metric completion of the equivalence classes $[w]_\otimes$ is required. The equalities guarantee that the functions are well-defined and $(W, 0_W, +_W, \cdot_W)$ forms a Hilbert space such that
\begin{eqnarray*}
\otimes : \Bilin{U}{V}{W}
\end{eqnarray*}
is bilinear, that is pointwise linear in each of its arguments.  Indeed, the operation $\otimes$ and the space $U \otimes V$ are constructed to be \emph{universal}: For every bilinear function $\diamond: \Bilin{U}{V}{T}$ there exists a unique linear function $\bi{\diamond}: \Hom{U \otimes V}{T}$ such that $\diamond = \bi{\diamond} \comp \otimes$.

Furthermore, we define the inner product
\begin{eqnarray*}
\odot : \Bilin{W}{W}{\Real}
\end{eqnarray*}
to be the unique bilinear function that satisfies
\begin{eqnarray*}
(u_1 \otimes v_1) \odot (u_2 \otimes v_2) & = & (u_1 \odot u_2) \cdot (v_1 \odot v_2).
\end{eqnarray*}

\section{Functions in combinatory form}
\label{combinatoryform.sec}

We provide a domain-specific language for specifying analytic functions on Hilbert spaces in combinatory form, that is in point-free notation.
In combinatory form, all subterms are closed functions; in particular, a subterm does not have implicit dependencies on an environment.  This facilitates formulation of a compositional differential calculus for calculating Fr\'echet derivatives.  

\subsection{Tensor contraction}

We provide a language constant for a single bilinear function.  It would be sufficient to provide the tensor product $\otimes$ as sole bilinear function since it is universal in the sense that all bilinear functions $f: \Bilin{U}{V}{W}$ factor into $f = \bi{f} \lcomp (\otimes)$ for a unique $\bi{f}: \Hom{U \otimes V}{W}$, the characteristic universal property of $\otimes$.
For reasons to become clear later, we provide \emph{tensor contraction}
\begin{eqnarray*}
\tensor & : & \Bilin{(W \otimes V)}{(V \otimes U)}{(W \otimes U)}
\end{eqnarray*}
instead.  It is defined as the unique bilinear function satisfying
\begin{eqnarray*}
(w \otimes v) \tensor (v' \otimes u) & = & (v \odot v') \cdot (w \otimes u)
\end{eqnarray*}

\subsection{Unitary operators}

We have a large number of useful \emph{natural unitary operators}; these are natural linear isomorphisms that are \emph{isometric}, i.e.~preserve norms.  We list a few of them here.
\[ \begin{array}{rcccl}
\bra{\_} & : & \Iso{V}{\Real \otimes V} & : & \ibra{\_} \\
\ket{\_} & : & \Iso{V}{V \otimes \Real} & : & \iket{\_} \\
\transpose{\_} & : & \Iso{(V \otimes W)}{(W \otimes V)} & : & \transpose{\_} \\
\assoc & : & \Iso{((U \otimes V) \otimes W)}{(U \otimes (V \otimes W))} & : & \iassoc \\
\distrib & : & \Iso{(\biprod{x \in X}{V_x}) \otimes W}{\biprod{x \in X}{(V_x} \otimes W)} & : & \idistrib \\
\zip & : & \Iso{(\biprod{x \in X}{V_x}) \oplus (\biprod{x \in X}{W_x})}{\biprod{x \in X}{(V_x \oplus W_x)}} & : & \unzip
\end{array} \]
They are defined by
\begin{eqnarray*}
\bra{v} & = & 1 \otimes v \\
\ibra{k \otimes v} & = & k \cdot v \\
\ket{v} & = & v \otimes 1 \\
\iket{v \otimes k} & = & k \cdot v \\
\transpose{(v \otimes w)} & = & w \otimes v \\
\assoc ((u \otimes v) \otimes w) & = & u \otimes (v \otimes w) \\
\iassoc (u \otimes (v \otimes w)) & = & (u \otimes v) \otimes w \\
\distrib ((\oplus_{x \in X} v_x) \otimes w) & = & \oplus_{x \in X} (v_x \otimes w) \\
(\zip(v,w))_x & = & (v_x, w_x)
\end{eqnarray*}
where $\oplus_{x \in X} v_x$ is notation for the element of $\oplus_{x \in X} V_x$ that maps $x$ to the value $v_x \in V_x$.
It turns out that the inverse of a unitary operator is also its adjoint; this will be useful later.

A derived isometric isomorphism is
$$\Iso{\coprod{X}{V} \otimes \coprod{Y}{W}}{\coprod{X \times Y}{(V \otimes W)}}$$
and in particular
$$\Iso{\prod{\seg{m}}{\Real} \otimes \prod{\seg{n}}{\Real}}{\prod{\seg{m} \times \seg{n}}{\Real}}.$$
In other words, all the elements of the tensor product of $\prod{\seg{m}}{\Real}$ and $\prod{\seg{n}}{\Real}$ can be 
represented by $m \times n$ matrices. Our construction of $\prod{\seg{m}}{\Real} \otimes \prod{\seg{n}}{\Real}$ using  
symbolic operators $0$, $\cdot$, $+$ and $\otimes$ provides more space efficient representations for low-rank matrices, however.
For example, every rank-1 $m \times n$ matrix corresponds to $v \otimes w$, its \emph{tensor decomposition}, for some $v \in \prod{\seg{m}}{\Real}, w \in \prod{\seg{n}}{\Real}$.  This term representation is of size $O(m + n)$ rather than requiring $m \cdot n$ entries in a matrix. (Note that a rank-1 matrix may have no $0$-entries.) Matrix/vector multiplication can be performed with only $n$ multiplications instead of $m \cdot n$ multiplications when using the matrix representation.   
  
The tensor and inner product operators are special cases of tensor contraction via the $\bra{\_}$ and $\ket{\_}$ unitary operators:
\begin{eqnarray*}
v \otimes w & = & \ket{v} \tensor \bra{w} \\
v_1\odot v_2 & = & \ibra{\bra{v_1} \tensor \ket{v_2}}
\end{eqnarray*}
Note that these are parsed as $(\ket{v}) \tensor (\bra{w})$ and $\ibra{(\bra{v}) \tensor (\ket{w})}$, respectively.

\subsection{Linear functions}

In addition to the unitary operators, the following are linear functions:
\[ \begin{array}{rcll}
(v \tensor) & : & \Hom{V \otimes U}{W \otimes U} & \mbox{if } v \in W \otimes V \\
(\tensor w) & : & \Hom{W \otimes V}{W \otimes U} & \mbox{if } w \in V \otimes U \\
0_{V,W} & : & \Hom{V}{W} & \\
\inj{y}{X} & : & \Hom{V_y}{\biprod{x \in X}{V_x}} & \mbox{if } y \in X \\
\proj{y}{X} & : & \Hom{\biprod{x \in X}{V_x}}{V_y} & \mbox{if } y \in X \\
\zipapp_{x \in X} f_x & : & \Hom{\biprod{x \in X}{V_x}}{\biprod{x \in X}{W_x}} & \mbox{if } f_x : \Hom{V_x}{W_x}\\
\Delta f & : & \Hom{\biprod{x \in X}{V_x}}{\biprod{x \in X}{W_x}} & \mbox{if } f : \biprod{x \in X}{(\Hom{V_x}{W_x})} \\
\langle f_x \rangle_{x \in X} & : & \Hom{V}{\biprod{x \in X}{W_x}} & \mbox{if } f_x : \Hom{V}{W_x} \\
\red_R & : & \Hom{\coprod{X}{V}}{\coprod{Y}{V}} & \mbox{if } R \subseteq X \times Y  \mbox{ is compact} \\
\prod{X}{f} & : & \Hom{\coprod{X}{U}}{\coprod{X}{V}} & \mbox{if } f : \Hom{U}{V} \\
\id_V & : & \Hom{V}{V} & \\
g \lcomp f & : & \Hom{U}{W} & \mbox{if } f : \Hom{U}{V}, g : \Hom{V}{W}
\end{array} \]

%

\subsection{Constant functions}

We have the \emph{constant functions}
\[ \begin{array}{rcll}
\const{w} & : & \Func{V}{W} & \mbox{if } w \in W
\end{array} \]
defined by $\const{w}(v) = w$.

\subsection{Primitive functions}

We furthermore assume we have named primitive functions $p_1, \ldots, p_n$ denoting analytic functions with associated derivative functions that are expressible as combinator expressions.
For example, for each $k \in \Int/\{0\}$ we have the function $\power{k}{\_} : \Func{\Real}{\Real}$ with associated derivative $\deriv{(\power{k}{\_})}(x) = ((k \cdot \power{k-1}{x}) \cdot)$.  Note the $\cdot$ at the end; it is there since the Fr\'echet derivative at $x$ is not a value from $\Real$, but an element of $\Hom{\Real}{\Real}$, which is isomorphic with, but not the same as, $\Real$.
Similarly, we have $\ln : \Func{\Real}{\Real}$ with associated $\deriv{\ln}(x) = (\power{-1}{x} \cdot)$; $\sin : \Func{\Real}{\Real}$ with
$\deriv{\sin}(x) = ((\cos x) \cdot)$; $\cos : \Func{\Real}{\Real}$ with $\deriv{\cos}(x) = ((-\sin x) \cdot)$ and so on. Note that $\ln$ is only
defined on $\Real_+ = \{ x \in \Real \mid x > 0 \}$ and is thus, in particular, not analytic on all of $\Real$.  

We defer the subtleties of handling partially defined and not-everywhere differentiable functions in this paper to future work and assume henceforth for simplicity that our primitive functions are analytic on their entire domain.

In practice almost all primitive functions are functions on scalars and returning scalars.  Primitive operators and functions on high-dimensional spaces are typically linear or bilinear.

\subsection{Function composition}

Every constant, linear, bilinear and primitive function constructed so far is an analytical function.

Finally we have sequential and parallel composition of analytical functions:
\[ \begin{array}{rcll}
g \comp f & : & \Func{U}{W} & \mbox{if } f : \Func{U}{V}, g : \Func{V}{W} \\
\zipapp_{x \in X} f_x & : & \Func{\biprod{x \in X}{V_x}}{\biprod{x \in X}{W_x}} & \mbox{if } f_x : \Func{V_x}{W_x} \mbox{ for all } x \in X, X \mbox{ finite}
\end{array} \]

\section{Fr\'echet differential calculus}
\label{differential-calculus}

Recall that $\deriv{f} : \Func{V}{(\Hom{V}{W})}$ denotes the Fr\'echet derivative of $f : \Func{V}{W}$. The linear function $\deriv{f}(v)$ is the tangent of $f$ at $v$.  We provide differentiation rules for functions in combinatory form.
\begin{theorem}
\label{differentiation-rules}
The following differentiation rules are valid for analytic functions on Hilbert spaces:
\begin{eqnarray}
\deriv{(g \comp f)}(v) & = & \deriv{g}(f(v)) \lcomp \deriv{f}(v) \label{chain-rule} \\
\deriv{\const{w}}(v) & = & 0 \label{constant-rule} \\
\deriv{h}(v) & = & h \qquad\qquad\qquad\qquad\qquad\!\!\! \mbox{if } h: \Hom{V}{W} \label{lin-rule} \\
\deriv{\diamond}(u, v) & = & (u \diamond) \lcomp \pi_2 + (\diamond v) \lcomp \pi_1 \qquad \mbox{if } \diamond: \Bilin{U}{V}{W} \label{bilin-rule} \\
\deriv{(\zipapp_{x \in X} f_x)}(v) & = & \Delta ((\zipapp_{x \in X} \deriv{f_x})(v)) \qquad\qquad\!\!\!\!\! \mbox{if } f_x: \Func{V_x}{W_x} \label{par-rule}
\end{eqnarray}
\end{theorem}
Rule~\ref{chain-rule} is the \emph{chain rule} for sequential composition.  It expresses that the derivatives of $g$ at $f(v)$ and of $f$ at $v$ are combined by composition $\lcomp$ of linear functions.  

Rules~\ref{constant-rule}, \ref{lin-rule} and \ref{bilin-rule} are for constant, linear and bilinear functions, respectively.  Note in particular Rule~\ref{bilin-rule}, the \emph{generalized product rule}. It is applicable to any bilinear function.  The derivative of any bilinear function can be written in terms of the function itself; we do not need access to its definition, only its name.  The same is true for linear functions; they are their own derivatives.  All we need to know is that a function is linear to differentiate it. We will see that adjoint differentiation, which underlies reverse-mnode AD, requires processing its definition, however. 

Finally, Rule~\ref{par-rule} is for differentiating \emph{parallel composition}.  It is worth looking at special cases of it. Let $X = \seg{2}$, that is
$\zipapp_{x \in \seg{2}} f_x = f_1 \times f_2 : \Func{V_1 \times V_2}{W_1 \times W_2}$.  We can calculate
\begin{eqnarray*}
\deriv{(f_1 \times f_2)}(v_1, v_2) & = & \Delta ((\deriv{f_1} \times \deriv{f_2})(v_1, v_2)) \\
& = & \Delta (\deriv{f_1}(v_1), \deriv{f_2}(v_2)) \\
& = & \deriv{f_1}(v_1) \times \deriv{f_2}(v_2)
\end{eqnarray*}
Let us consider $\prod{X}{f} = \zipapp_{x \in X} f$ where $f \in \Func{V}{W}$.
\begin{eqnarray*}
\deriv{(\prod{X}{f})}(v) & = & \deriv{(\zipapp_{x \in X} f)}(v) \\
& = & \Delta ((\zipapp_{x \in X} \deriv{f})(v)) \\
& = & \Delta (\prod{X}{\deriv{f}}(v))
\end{eqnarray*}
In words, to differentiate $\prod{X}{f}$ at value $v \in \coprod{X}{V}$, we need to compute the derivative of $f$ at each element $v_x$ of $\coprod{X}{V}$.
This yields
an element of $\coprod{X}{(\Hom{V}{W})}$; finally, $\Delta$ gathers these component-wise derivatives into a single derivative.

\section{Affine interpretation}
\label{affine-interpretation.sec}

A function $h: \Func{V}{W}$ is \emph{affine} if it is the sum of a constant and a linear function, that is
$$h(v) = w + g(v)$$  for some $w \in W$ and $g \in \Hom{V}{W}$.  Note that $w$ and $g$ are uniquely determined by $h$. We call them the \emph{constant} and \emph{linear} component of $h$, respectively, and write $h \in \Aff{V}{W}$ if $h$ is affine. 

We say that $g : \Aff{V}{W}$ is the \emph{affine approximation of $f : \Func{V}{W}$ at $v \in V$} and write $f(v) \approx g$ if
\begin{equation*}
\lim_{||\dv||_V \rightarrow 0} \frac{||f(v + \dv) - g(\dv)||_W}{||\dv||_V} = 0 
\end{equation*}

\begin{proposition}
A function has at most one affine approximation at $v$, written $\aff{f}(v)$, where $\aff{f}(v)(\dv) = f(v) + \deriv{f}(v)(\dv)$.
\end{proposition}
Thinking about differentiation in terms of computing affine approximations is useful since computing derivatives compositionally requires computing a function's value paired with its derivative \cite{elliott2018simple}.  The components of the affine approximation of a function in combinatory form can be computed by structural recursion. See Figure~\ref{affine-interpretation}.



\begin{figure}
\[ \begin{array}{rcll}
\aff{(g \comp f)}(x) & = & \Let (\fx, \fpx) = \aff{f}(x) \In \\
                     & & \Let (\gfx, \gpfx) = \aff{g}(\fx) \In \\
                     & & \qquad (\gfx, \gpfx \lcomp \fpx) \\
\aff{\const{w}}(x) & = & (w, 0) \\
\aff{h}(x) & = & (h(x), h) & \mbox{if } h: \Hom{V}{W} \\
\aff{\diamond}(x) & = & \ULet (u, v) = x \In \\
                  & & \qquad (u \diamond v, (u \diamond) \lcomp \pi_2 + (\diamond v) \lcomp \pi_1) & \mbox{if } \diamond: \Bilin{U}{V}{W} \\
\aff{(\zipapp_{y \in Y} f_y)}(x) & = & \Let (w, d) = \unzip((\zipapp_{y \in Y} (\lambda x. \aff{f_y}(x)))(x)) \In \\
                  & & \qquad (w, \Delta(d)) \qquad\mbox{if } f_y: \Func{V_y}{W_y} 
\end{array} \]
\caption{Affine interpretation of functions in combinatory form. See Section~\ref{symbolic-differentiation} for an explanation of the underlined $\ULet$.}
\label{affine-interpretation}
\end{figure}

The parallel composition rule specializes to tuples and copowers as follows:
\[\begin{array}{rcl}
\aff{(f_1 \times f_2)}(x_1, x_2) & = & \Let (\fx_1, \fpx_1) = \aff{f_1}(x_1), (\fx_2, \fpx_2) = \aff{f_2}(x_2) \In ((\fx_1, \fx_2), (\fpx_1, \fpx_2)) \\
\aff{\prod{X}{f}}(v) & = & \Let (w, d) = \unzip(\prod{X}{\aff{f}}(v)) \In (w, \Delta(d))
\end{array} \]
Note that unzipping the outputs of each component is the price we pay for separating the collective output into a value and a derivative component.

\begin{theorem} Assume $\aff{p}(v) = (p(v), \deriv{p}(v))$ for all primitive functions.  Then
$\aff{f}(v) = (f(v), \deriv{f}(v))$ for all functions in combinatory form.
\end{theorem}

\subsection{Automatic differentation}

The affine approximation rules give rise to an interpreter $$\eval{\_}: \Term{\Func{V}{W}} \rightarrow V \rightarrow \Term{W \times (\Hom{V}{W})}$$
where $\Term{\Func{V}{W}}$ is a language for representing functions in combinatory form, including $\Term{\Hom{V}{W}}$ as a (sub)language for representing linear functions in combinatory form: Just replace $\aff{t}$ in Figure~\ref{affine-interpretation} by $\eval{t}$.  When applied to a combinatory term $t$ denoting $f$ and a concrete value $v$, it returns a term containing the value $w = f(v)$ and a combinatory representation $t'$ denoting the derivative $\deriv{f}(v)$.  This term $t'$ can be optimized using the rules of linear and tensor algebra prior to applying an
interpreter $\evalz{\_} : \Term{\Hom{V}{W}} \rightarrow V \rightarrow W$ to $t'$ and an input differential value $\dv$, which yields the output differential $\deriv{f}(v)(\dv)$.

Behaviorally this corresponds to forward-mode automatic differentiation (AD), but is essentially different.
In \emph{elemental} and \emph{tensor-based} forward-mode AD \cite{griewankAutomaticDifferentiation,abadi2016tensorflow}
we have an interpreter $\evalf{\_}$ for a term $t : \Term{\Func{\biprod{i \in \seg{n}}{(\prod{\seg{2}}{V_i})}}{\biprod{j \in \seg{m}}{(\prod{\seg{2}}{V_j})}}}$ representing $f$ such that
\begin{eqnarray*}
\evalf{t} & : & \Func{\biprod{i \in \seg{n}}{(\prod{\seg{2}}{V_i})}}{\biprod{j \in \seg{m}}{(\prod{\seg{2}}{V_j})}}.
\end{eqnarray*}
requires both values and associated differentials as inputs at the same time.
It computes
\begin{eqnarray*}
\evalf{t}((v_1, \dv_1), \ldots, (v_n, \dv_n)) & = & ((y_1, \dy_1), \ldots, (y_m, \dy_m)).
\end{eqnarray*}
where $(y_1, \ldots, y_m) = f(v_1, \ldots, v_n)$ and
$(\dy_1, \ldots, \dy_m) = \deriv{f}(v_1, \ldots, v_n)(\dv_1, \ldots, \dv_n)$.

The type of $\evalf{t}{}$ camouflages that the output values $y_1, \ldots, y_n$ do not depend on the second components $\dv_1, \ldots, \dv_n$,
and that for fixed $v_1, \ldots, v_n$ the $\dy_1, \ldots, \dy_m$ are linear functions of $\dv_1, \ldots, \dv_n$. Note, in particular, that both values and differentials must be provided before execution can start.

In our formulation $\eval{t}$ requires no input differentials to run the code, only the input values $v_1, \ldots, v_n$, which
manifests that the output values do not depend on any differentials.
Furthermore, the derivative is returned as a term in a language that guarantees that it denotes a linear function.  
%
%

\subsection{Symbolic differentiation}
\label{symbolic-differentiation}

The affine approximation rules are carefully written to facilitate \emph{symbolic differentiation} by applying $\eval{t}$ to a symbolic variable $x$. This amounts to \emph{specializing} the code of $\eval{\_}$ to the concrete $t$ by partial evaluation.

The $\Let$-expressions without underlining can be eliminated by substitution, that is rewriting $\Let (u, v) = (f, g) \In g$ to $g[f/u, g/v]$ during partial evaluation.  Since the let-bound variables have single occurrences the size of the expression does not grow.  The $\ULet$-expression for bilinear functions should not be eliminated, however, since its let-bound variables are used twice.  Substituting them would cause \emph{expression swell}.  Conversely, not substituting them avoids expression swell: The size of the symbolically differentiated expression is linear in the size of the input expression.  Retaining $\ULet$ in the output is the reason for having 
$$\eval{\_}: \Term{\Func{V}{W}} \rightarrow V \rightarrow \Term{W \times (\Hom{V}{W})}$$
rather than
$$\eval{\_}: \Term{\Func{V}{W}} \rightarrow V \rightarrow (W \times \Term{\Hom{V}{W}}).$$
This supports and generalizes to non-elemental symbolic differentiation that expression swell is a myth  \cite{laue2019equivalence}: Retaining sharing
when applying the (generalized) product rule is both necessary and sufficient to avoid it.

\section{Adjoints}
\label{adjoints}

The \emph{dual vector space} $\dual{V}$ of vector space $V$ is the vector space of \emph{linear functionals}, also called \emph{covectors}, $\Hom{V}{\Real}$ where $\Real$ is the underlying field of $V$.
By the Riesz representation theorem, the inner product induces an isomorphism $\dualvec$ defined by
\begin{eqnarray*}
\dualvec & : & \Hom{V}{V^*} \\
\dualvec(v) & = & (\odot v).
\end{eqnarray*}
In particular, $\inverse{\dualvec}(\odot v) = v$, and $v$ and $(\odot v)$ are called \emph{duals} of each other.\footnote{For finite index sets the constructible Hilbert spaces are finite-dimensional.  Note that the Riesz representation theorem also holds for infinite-dimensional Hilbert spaces.}

Some applications require computing the dual of a covector.  For example, given a \emph{scalar} function $f : \Func{V}{\Real}$, the \emph{gradient} $\grad f : \Func{V}{V}$ is defined by
$$\grad f (v) = \inverse{\dualvec}(\deriv{f}(v)).$$
If we implement covectors as functions that can only be applied, the only way of implementing $\inverse{\dualvec}$ is by applying it to each of the base vectors of $V$, which is problematic if $V$ is of high dimension, say a million or a billion.

Similarly, sometimes we may want to implement the \emph{transpose}
\begin{eqnarray*}
\adjoint{f} & : & \Hom{\dual{W}}{\dual{V}} \\
\adjoint{f} & = & (\lcomp f)
\end{eqnarray*}
of $f: \Hom{V}{W}$.
The transpose is the \emph{continuation-passing style} version of $f$ where a linear continuation is passed as the first argument.  To wit, we have
$\adjoint{f}(\kappa)(v) = \kappa(f(v))$
where $\kappa$ is the continuation.

A general idea permeating mathematical and computer science applications of linear algebra is representing a covector by its dual vector $v$ with an indication that it represents $(\odot v)$ (``I am contravariant''), not $v$ itself.

We would thus like to find a linear function $\adj{f} : \Hom{W}{V}$ that implements the transpose $\adjoint{f}$ by using ordinary vectors rather than covectors to represent linear continuations; that is, it should be the case that $\adj{f}(w) = v$ whenever $\adjoint{f}(\odot w) = (\odot v)$.
\begin{definition}
$\adj{f}: \Hom{W}{V}$ is the \emph{adjoint} of $f: \Hom{V}{W}$ if
$$\adjoint{f}(\odot w) = (\odot v) \Leftrightarrow \adj{f}(w) = v$$
\end{definition}
By the Riesz representation theorem we immediately have that
\begin{proposition}
$\adj{f}$ exists and is unique for Hilbert spaces.
\end{proposition}
The defining property of an adjoint can be restated as the familiar property where $f$ is pushed from one argument to the other argument of the inner product.
\begin{proposition}
$\adj{f} : \Hom{W}{V}$ is the adjoint of $f: \Hom{V}{W}$ if and only if $f(v) \odot w = v \odot \adj{f}(w)$
for all $v \in V, w \in W$.
\end{proposition}
We can implement $\inverse{\dualvec}$ using the adjoint:
\begin{proposition}
Let $f : \Hom{V}{\Real}$, that is $f \in \dual{V}$. Then $\inverse{\dualvec}(f) = \adj{f}(1)$ and thus 
$\grad f (v) = \adj{(\deriv{f}(v))}(1)$.
\end{proposition}
Linear functions are built from other linear functions.  We provide general rules for calculating adjoints symbolically of linear functions in combinatory form.  

\subsection{Adjoint calculation}

Adjoints can be calculated symbolically for linear functions in combinatory form.
\begin{theorem}
\label{adjoint-rules}
Let $X, Y$ be finite sets$, R \subseteq X \times Y$, and $\transpose{R} = \{ (y, x) \mid (x, y) \in R \}$. Then:
\begin{eqnarray*}
\adj{\id} & = & \id \\
\adj{(g \lcomp f)} & = & \adj{f} \lcomp \adj{g} \\
\adj{0} & = & 0 \\
\adj{(v \tensor)} & = & (\transpose{v} \tensor) \\
\adj{(\tensor w)} & = & (\tensor \transpose{w}) \\
\adj{(\inj{x}{X})} & = & \proj{x}{X} \\
\adj{(\zipapp_{x \in X}{f_x})}& = & \zipapp_{x \in X}{\adj{f_x}} \\
\adj{\red_R} & = & \red_{\transpose{R}}
\end{eqnarray*}
Furthermore, the inverses of unitary operators are also their adjoints.
\end{theorem}
These rules are not accidentally symmetric; indeed the above language of linear functions has been \emph{designed} to yield these symmetries, where each construct has an adjoint construct.  This is the reason for using tensor contraction $\tensor$ instead of $\otimes$ as the primitive universal bilinear function.

The adjoint of a partially applied tensor contraction $(v \tensor)$ is the \emph{transpose}\footnote{Not to be confused with the transpose of a linear function, which it is related, but different.} $\transpose{v}$ of $v$.
Recall that $\transpose{\_} : \Hom{V \otimes W}{W \otimes V}$ swaps components of formal tensor product sums.

Note also that the adjoint of $\red_R$ is $\red_{R^T}$; in particular, if $R$ is a function, $R^T$ is generally not a function.  Allowing for $R$ to be a relation rather than restricting it to be a function makes expressing its adjoint in terms of $\red$ possible.

The adjoints for other operations can be derived from their definitions in terms of these primitive functions and constructs.
For example, we can derive
\begin{eqnarray*}
\adj{(k \cdot)} & = & (k \cdot) \\
\adj{(\cdot v)} & = & (\odot v)
\end{eqnarray*}

\subsection{Adjoint differentiation}
\label{sec:adjdiff}

The \emph{adjoint derivative} of $f : \Func{V}{W}$ at $v \in V$ is $\adj{(\deriv{f}(v))}$.  We can compute it by employing our affine interpreter $\eval{f}$; applying it to $v$; extracting the term representing $\deriv{f}(v)$;
applying the adjoint calculation rules of Theorem~\ref{adjoint-rules} to calculate a term representating $\adj{(\deriv{f}(v))}$; and finally applying the 
derived adjoint to output differentials to compute input differentials.\footnote{When used in the adjoint direction, from output to input, the variables containing differentials are often called adjoint variables.}
Alternatively, we can construct the adjoint derivative during affine interpretation; see Figure~\ref{adjoint-affine-interpretation}.

\begin{figure}
\[ \begin{array}{rcll}
\evalreverse{(g \comp f)}(x) & = & \Let (\fx, \fpxa) = \evalreverse{f}(x) \In \\
                     & & \Let (\gfx, \gpfxa) = \evalreverse{g}(\fx) \In \\
                     & & \qquad (\gfx, \fpxa \lcomp \gpfxa ) \\
\evalreverse{\const{w}}(x) & = & (w, 0) \\
\evalreverse{h}(x) & = & (h(x), \adj{h}) & \mbox{if } h: \Hom{V}{W} \\
\evalreverse{\diamond}(x) & = & \ULet (u, v) = x \In \\
                  & & \qquad (u \diamond v, \inj{2}{\seg{2}} \lcomp \adj{(u \diamond)} + \inj{1}{\seg{2}} \lcomp \adj{(\diamond v)}) & \mbox{if } \diamond: \Bilin{U}{V}{W} \\
\evalreverse{(\zipapp_{y \in Y} f_y)}(x) & = & \Let (w, d) = \unzip((\zipapp_{y \in Y} (\lambda x. \evalreverse{f_y}(x)))(x)) \In \\
                  & & \qquad (w, \Delta(d)) & \mbox{if } f_y: \Func{V_y}{W_y}
\end{array} \]
\caption{Adjoint affine interpretation of functions in combinatory form}
\label{adjoint-affine-interpretation}
\end{figure}

This provides us with a method behaviorally equivalent to reverse-mode automatic differentiation.
In the first phase, the value of $f$ at $v$ and the term of $\adj{(\deriv{f}(v))}$, which
includes all---and only--- the relevant intermediate results of $f(v)$ are computed by $\evalreverse{f}$ from $v$ alone.  Only in the second phase, the output term representing $\adj{(f(v))}$ is interpreted as a function by applying it to output differentials.  After the first phase, the adjoint derivative can be optimized using algebraic simplications, and it can be compiled for efficient data parallel execution on a GPU.

Going one step further, the first phase can be done symbolically, which amounts to a specialization of the adjoint affine interpreter to the particular source code for $f$.  The result of doing so can be compiled for efficient data parallel execution before the values of $v$ and output differential $\dy$ are available.

\section{Applications}
\label{applications.sec}

We illustrate combinatory differentiation by applying it to neural networks.  Additional examples showing the application of equational reasoning to derive derivatives (sic!) can be found in Appendix~\ref{applications}.

\begin{example}
\label{nn-example}
A $k$-layer neural network $N_k$ consists of a composition of $k$
layers with the $i$-th layer given by
\[
g_i(x_i,W_i,b_i) = \prod{\seg{m}_i}{h_i}  \, (W_i \mmult x_i + b_i),
\]
where $W_i \in \prod{\seg{m}_i \times \seg{m}_{i-1}}{\mathbb{R}}$, $b_i \in \prod{\seg{m}_i}{\mathbb{R}}$, and $x_i \in \prod{\seg{m}_{i-1}}{\mathbb{R}}$
along with a loss function $l(v,y) : \prod{\seg{m}_k}{\mathbb{R}} \rightarrow \mathbb{R}$:
\begin{align*}
&N_k : \prod{\seg{m}_0}{\mathbb{R}}
      \times \prod{\seg{m}_1 \times \seg{m}_0}{\mathbb{R}}
      \times \prod{\seg{m}_1}{\mathbb{R}}
      \times \prod{\seg{m}_2 \times \seg{m}_1}{\mathbb{R}}
      \times \prod{\seg{m}_2}{\mathbb{R}}
      \times \cdots
      \times \prod{\seg{m}_k \times \seg{m}_{k-1}}{\mathbb{R}}
      \times \prod{\seg{m}_k}{\mathbb{R}}
      \times \prod{\seg{m}_k}{\mathbb{R}}
      \rightarrow \mathbb{R} \\
&N_k(x,W_1,b_1,W_2,b_2,\dots,W_k,b_k,y) = l(g_k(W_k,b_k(\cdots(g_{2}(W_2,b_2(g_1(W_1,b_1,x)))))), y).
\end{align*}
For simplicity, we use $l(v,y) = (v - y) \odot (v - y)$ for the loss function.
In point-free form, the $i$-th layer of the network must propagate the
inputs for all subsequent layers; $g_i$ and $l$ in point-free form are
\begin{align*}
g_i &= \langle\prod{\seg{m}_i}{h_i}~\comp~((\mmult) \circ \langle \proj{2}{\seg{n}_i}, \proj{1}{\seg{n}_i}\rangle +\proj{3}{\seg{n}_i})
        , \proj{4}{\seg{n}_i}, \dots, \proj{n_i}{\seg{n}_i} \rangle, \\
l   &= (\odot) \circ \dup \circ (\proj{1}{\seg{2}} - \proj{2}{\seg{2}}),
\end{align*}
where $\seg{n}_i = 2(k+2-i)$. Hence, the entire network is constructed as
\[
N_k = l \circ g_k \circ \cdots g_2 \circ g_1.
\]
Applying the differentiation rules of Theorem \ref{differentiation-rules}, we differentiate $g_i$ and $l$
\begin{align*}
&g_i'(x_i,W_i,b_i,\dots,W_k,b_k,y) \\
 \quad &= \mbox{\{definition of $g_i$\}} \\
  &\quad\langle\prod{\seg{m}_i}{h_i}~\comp~((\mmult) \circ \langle \proj{2}{\seg{n}_i}, \proj{1}{\seg{n}_i}\rangle +\proj{3}{\seg{n}_i})
        , \proj{4}{\seg{n}_i}, \dots, \proj{n_i}{\seg{n}_i} \rangle' (x_i,W_i,b_i,\dots,W_k,b_k,y)\\
 \quad &= \mbox{\{by Rules \ref{chain-rule}, \ref{lin-rule}, and \ref{par-rule}\}} \\
          &\quad\langle(\prod{\seg{m}_i}{h_i})'(W_i \mmult x_i + b_i)
          ~\lcomp~((\mmult)' (\langle W_i, x_i\rangle) \lcomp \langle \proj{2}{\seg{n}_i}, \proj{1}{\seg{n}_i}\rangle
           +\proj{3}{\seg{n}_i}),\proj{4}{\seg{n}_i}, \dots,  \proj{n_i}{\seg{n}_i}\rangle\\
 \quad &= \mbox{\{by Rules \ref{par-rule} and \ref{bilin-rule} \}} \\
          &\quad\langle \Delta ({\prod{\seg{m}_i}{h_i'}} (W_i \mmult x_i + b_i))
          ~\lcomp~(((W_i \mmult) \lcomp \proj{2}{\seg{2}} + (\mmult x_i) \lcomp \proj{1}{\seg{2}}) \lcomp \langle \proj{2}{\seg{n}_i}, \proj{1}{\seg{n}_i}\rangle
           +\proj{3}{\seg{n}_i}),\proj{4}{\seg{n}_i}, \dots,  \proj{n_i}{\seg{n}_i}\rangle,\\\\
&l'(v,y) \\
\quad &= \mbox{\{definition of $l$\}} \\
 &\quad ((\odot) \circ \dup \circ (\proj{1}{\seg{2}} - \proj{2}{\seg{2}}))'(v,y) \\
\quad &= \mbox{\{by Rule \ref{chain-rule}\}} \\
 &\quad (\odot)'(v-y, v-y)  \lcomp (\dup \circ (\proj{1}{\seg{2}} - \proj{2}{\seg{2}}))'(v,y) \\
\quad &= \mbox{\{by Rules \ref{chain-rule}, \ref{bilin-rule}, and \ref{lin-rule}\}} \\
 &\quad (((v - y) \odot) \lcomp \proj{2}{\seg{2}} + (\odot (v-y)) \lcomp \proj{1}{\seg{1}})  \lcomp (\dup \lcomp (\proj{1}{\seg{2}} - \proj{2}{\seg{2}})).
\end{align*}
Repeated application of Rule \ref{chain-rule} now yields the entire differentiated network
\begin{align*}
N_k'(x,W_1,b_1,W_2,b_2,\dots,W_k,b_k,y) &= l'((g_k \comp \cdots \comp g_1)(x,W_1,b_1,W_2,b_2,\dots,W_k,b_k,y)) \\
                                       &\qquad \lcomp ~g_k'((g_{k-1} \comp \cdots \comp g_1)(x,W_1,b_1,W_2,b_2,\dots,W_{k},b_{k},y)) \\
                                       & \qquad  \lcomp \cdots \lcomp ~ g_1'(x,W_1,b_1,W_2,b_2,\dots,W_{k},b_{k},y).
\end{align*}
Straight-forward application of Theorem \ref{adjoint-rules} may subsequently be used to obtain the adjoint of $N_k'$, $\adj{(N_k'(x,W_1,b_1,W_2,b_2,\dots,W_k,b_k,y))} : 
      \Hom{\mathbb{R}}{ 
      \prod{\seg{m}_0}{\mathbb{R}}
      \times \prod{\seg{m}_1 \times \seg{m}_0}{\mathbb{R}}
      \times \cdots
      \times \prod{\seg{m}_k}{\mathbb{R}}}$.

\end{example}

\section{Discussion}
\label{discussion}

We have provided a functional-analysis based compositional framework for differentiation and adjoint differentiation that encompasses both symbolic and automatic differentiation.  It highlights that, very generally, adjoint differentiation is the combination of symbolic Fr\'echet differentiation and symbolic calculation of adjoints over Hilbert spaces, where both derivatives and adjoints retain the data parallelism in their input functions.

\paragraph{Why Hilbert spaces?}  A Hilbert space is a vector space that is equipped with an inner product $\odot$ and is metrically complete.  The inner product is crucial: it establishes an isomorphism with the dual space such that ordinary first-order vectors can be used to represent linear functionals rather than having to code these as procedures in a programming language.  This representation trick
is the essence of adjoints, which run linear functions in reverse, from output differential to input differential and, in particular, compute gradients as the input differentials resulting from  a single evaluation of the adjoint derivative to the output differential $1$.

The metric completness is, in some sense, irrelevant: it only pops up in the definition of Fr\'echet derivative and checking that it constitutes a valid model of the differentiation rules.  


\paragraph{Why symbolic tensor products?} We could define the tensor product of $\prod{\seg{m}}{\Real}$ and $\prod{\seg{n}}{\Real}$ to
be the matrix space $\prod{\seg{m} \times \seg{n}}{\Real}$, but matrices as data structures for derivatives are too inefficient for large values of $m, n$.
Symbolic tensor decompositions can provide more efficient representations \cite{gelss2017tensor}.

For example, Griewank \cite{griewank2012invented} gives $f: \Func{\prod{\seg{n}}{\Real}}{\prod{\seg{m}}{\Real}}$ defined by
\vspace{-0.1cm}
$$f(x) = b \sin (a^T x)$$
\vspace{-0.1cm}
with $a \in \prod{\seg{n}}{\Real}$ and $b \in \prod{\seg{n}}{\Real}$ as an example where computing the Jacobian derivative of $f$ at $x_0$ requires $m \cdot n$ multiplications, whereas the original function requires only $n + m$ multiplications.
But this is only due to insisting on representing the derivative as a Jacobian matrix.  
Translated into combinatory form and employing our differentiation rules we arrive at a corresponding representation of the matrix as
\vspace{-0.1cm}
$$c  \cdot (b \otimes a)$$
\vspace{-0.1cm} 
where $c = \cos(a^T x_0)$.  Note that this is the output: itt uses symbolic scalar product and tensor product operators.  This representation can be computed using only $n$ scalar multiplications.  Furthermore, applying it to  
a vector $\dx \in \prod{\seg{n}}{\Real}$ produces the term $d \cdot b$ where $d = c \cdot (a \odot \dx)$, which requires only $n+1$ multiplications.
Using the matrix equivalent of $c  \cdot (b \otimes a)$ takes $m \cdot n$ multiplications.  

\subsection{Related work}


The origin of symbolic differentiation using electronic computers for functions on scalar variables dates back to the 1950s \cite{kahrimanian1953analytical}.
Forward-mode AD for scalar variables 
was discovered independently by a number of
researchers in the 1950s and 1960s \cite{griewank2008evaluating}.
The history of reverse-mode AD dates back to the early 1970s and 
is surveyed by Griewank \cite{griewank2012invented}.
Linnainmaa \cite{linnainmaa1970,linnainmaaTaylorExpansionAccumulated1976} observed early on that reverse-mode AD on scalar variables 
consists of building a computation graph \cite{bauer1974computational} and 
then reversing its dependency arrows, which is tantamount to
transposing sparse matrices in a sequential composition of matrix 
multiplications.  
This observation has since been made repeatedly in both elemental and tensor 
settings. 

Derivatives of functions on scalar variables are conventionally represented
by Jacobian matrices.  Computing the matrix with a minimum number of steps
is NP-hard, however \cite{naumann2008optimal}.  As we have shown, 
matrices are often not even a good data structure for derivatives, however.
As we have shown, they can be represented more compactly and efficiently 
using a combinatory language for linear functions, including symbolic 
operators for scalar multiplication, addition and tensor product \cite{mathiesen2016,hkm2022}.

 

Functional languages have served well for exploring AD techniques both
as a host language for capturing AD techniques
\cite{karczmarczukFunctionalDifferentiationComputer1998,
  karczmarczukFunctionalCodingDifferential1999}, and as the language
under investigation, featuring, for instance, multi-variate functions,
higher-dimensional data, higher-order functions, higher-degree
differentiation (e.g., through a lazy infinite tower of derivatives)
\cite{10.1145/1596550.1596579,esben2012}, and even differentiation of
formal languages \cite{elliott2021symbolic}.  Whereas many of the
above-mentioned features are well-suited for forward-mode AD (no
memoization of primal values is needed), capturing the essence of reverse-mode AD
has proven difficult.  We believe this is 
due to using $\lambda$-calculus formulation \cite{pearlmutterReversemodeADFunctional2008} rather than
a combinatory formulation, representing the ``tape'' 
as (the code of) a function or procedure \cite{elliott2018simple,wangDemystifyingDifferentiableProgramming2019} and/or
employing matrices to represent linear functions instead of asymptotically more compact and efficient data structures
made possible by symbolic tensor products and useful constants (identity, projections and injections). 

Reverse-mode AD for higher-order languages has been studied by Mazza \cite{mazzaAutomaticDifferentiationPCF2021} and
on capturing reverse-mode AD by
building a library for functional representation general differentiation based on the
specifications of the functionality . 
Our work follows Elliott's \cite{elliott2018simple} lead:\footnote{
While inspired by Elliott's elegant presentation of Fr\'echet derivatives in a Haskell framework \cite{10.1145/1596550.1596579}, our work on functional-analysis based AD started in 2015 and developed independently of Elliott's work, but 
has so far remained unpublished except for a presentation at a Workshop in
honor of Tom Reps's 60th birthday in 2016 \cite{henglein2016b}.} It is also based on adjoint affine interpretation, 
but additionally employs specialized and efficient representations for linear functions, supports 
sums, tensor products and copowers, avoids expression swell, identifies and exploits general differentiation rules for bilinear operators, and supports relational reduction and parallel composition. 


Other work has focused on exploring how AD techniques can be applied
in a functional parallel setting while preserving the parallel
properties of functions, also in the differentiated code
\cite{paszke2021getting, 10.1145/3471873.3472975}.  
Recent work \cite{schenck2022ad} on reverse- and forward-mode AD operators embedded in Futhark \cite{futhark} has 
shown that even nested parallelism can be handled in reverse mode effectively with
excellent GPU-utilization and performance.  
Our combinatory language features powerful parallel operations and general differentiation and adjoint rules
that retain semantic data parallelism, but does not devise a general implementation method for compact representation 
of relations and efficient parallel implementation of relational reduction.  This is future work.

An approach to combinatory differentiation based on category theory rather than linear algebra is \emph{differential categories}~\cite{blutecockettseelydiffcat2006,cockettJSPLonlyonediff2017} and its many variations (e.g., \cite{blutecockettseely2009cartesiandiffcat, cockettcruttwellgallagherdiffrestcat2011, cockettReverseDerivativeCategories2019}). In brief, a differential category is an additive monoidal category with a differential combinator and a modality allowing differentiable morphisms to be identified by their signature. The variation most closely resembling the one presented here is that of \emph{reverse derivative categories}~\cite{cockettReverseDerivativeCategories2019}, as they can be thought of as categories of smooth maps equipped with the ability to take adjoints (a ``dagger''). This approach is ultimately closer to symbolic differentiation rather than AD, though it could be interesting to integrate our approach into a notion of differential category with a distinction between semantic and syntactic data (see also \cite{danvyTDPE1996}).

The categorical semantics of both forward and reverse mode AD with higher types was recently given a unified treatment in \cite{vakarreversead2021}, with models based on so-called \emph{biadditive} categories: indexed categories with biproducts at each index, preserved by reindexing. Interestingly, when applying the Grothendieck construction $\int (-)$ to a biadditive category $\mathbf{C}$, the resulting fibred category $\int \mathbf{C}$ describes forward mode AD, while its dual $\int \mathbf{C}^\mathrm{op}$ describes reverse mode AD. This highlights the formal connection between duality (via adjoints) and reverse mode AD, as also argued in \cite{cockettReverseDerivativeCategories2019} and in Section~\ref{sec:adjdiff}.


On the practical side, a variety of systems provide tooling for
automatically differentiating source code. These
tools include (but are far from limited to) Python tools such as
Autograd \cite{maclaurin2016phd}, JAX
\cite{jax2018github,NEURIPS202083d3d4b6}, C/C++ tools such as Adept
\cite{Hogan2014FRM}, ADOL-C \cite{Griewank1996AAC} and Tapenade \cite{10.1145/2450153.2450158}, DiffSharp for
F\# \cite{adsurvey}, tools for MATLAB
\cite{Bischof2002CST,doi:10.1137/080743627}, Julia
\cite{innesDonUnrollAdjoint2019}, FutharkAD \cite{schenck2022ad} and even tools for the LLVM IR \cite{10.1145/3458817.3476165}. Most of these tools feature both
forward-mode and backward-mode AD and are therefore applicable for a
variety of domains and applications, including physics simulation
\cite{NEURIPS202083d3d4b6}, finance
\cite{esben2012,Bischof2002ADf,Giles2005SmokingAF}
and economics \cite{Tadjouddine2009ADA}. AD has also received renewed
attention due to its application to deep learning, where backpropagation is reverse-mode AD for scalar functions, as shown in Example \ref{nn-example}. AD techniques have therefore been incorporated, either directly or indirectly (through
library APIs), into most of the major general machine learning
frameworks, including Caffe \cite{10.1145/2647868.2654889}, TensorFlow
\cite{abadi2016tensorflow}, and PyTorch
\cite{paszkeAutomaticDifferentiationPyTorch2017}. For a general
overview, consult \cite{adsurvey}.  Work has also been done at
benchmarking many of the commonly used AD tools
\cite{srajer2016benchmark}.

\paragraph{Acknowledgements.}
\begin{small}
This work was made possible by Independent Research Fund Denmark grants \emph{FUTHARK: Functional Technology for High-performance Architectures}, \emph{Deep Probabilistic Programming for Protein Structure Prediction (DPP)}, and DFF--International Postdoc 0131-00025B.  We would like to thank Gabriele Keller, Ken Friis Larsen and Dimitrios Vytionitis for collaborative discussions over the last six years that have greatly helped in developing the foundations of combinatory differentiation and our colleagues on FUTHARK and DPP, in particular Cosmin Oancea, Troels Henriksen, Thomas Hamelryck and Ola R\o nning.  Furthermore, the second author would like to thank Conal Elliott
for stimulating exchanges on AD in the period he was working on his ICFP 2018 paper \cite{elliott2018simple}.  We greatly appreciate and thank the three anonymous referees for their recommendations. 
\end{small}

\bibliographystyle{eptcs}
\bibliography{p}

\appendix

\section{Derivatives}
\label{derivatives.sec}

%
Informally, the \emph{derivative} of a function $f$ at a particular input value $x$ is a mathematical object that describes how infinitesimal changes $\dx$ to $x$ incur changes $\dy$ to the result $y = f(x)$ of $f$ at $x$.
There are multiple notions of increasing generality and abstraction in mathematics that make ``describing'', ``infinitesimal'' and ``changes'' precise.

\subsection{Leibniz derivative}

For a scalar function of one (scalar) variable $f: \Real \rightarrow \Real$, the \emph{Leibniz derivative of $f$ at $x$} is
the number $a \in \Real$ that satisfies $$f(x + \dx) \approx  f(x) + a \mult \dx$$ where $\mult$ is multiplication on $\Real$ and $\approx$ expresses that the error on the right-hand side vanishes as $\dx$ becomes infinitesimally small.
Specifically, $a$ is the derivative of $f$ at $x$ if
$$\lim_{|\dx| \rightarrow 0} \frac{|f(x + \dx) - (f(x) + a \mult \dx)|}{|\dx|} = 0.$$
For example, for $f(x) = x^2$ we have that $8$ is the derivative of $f$ at $4$, and $14$ is the derivative of $f$ at $7$.

The \emph{Leibniz derivative of $f$} is the function $\deriv{f}$ that maps $x$ to the derivative of $f$ at $x$.
For example, for $f(x) = x^2$ we have $\deriv{f}(x) = 2 \mult x$.

\subsection{Jacobi derivative}

For a vector-valued function $f: \prod{\seg{n}}{\Real} \rightarrow \prod{\seg{m}}{\Real}$, the \emph{Jacobi derivative of $f$ at $v$} is
the $m \times n$-matrix $M$ that satisfies
$$f(v + \dv) \approx f(v) + M \mmult \dv.$$
Here $\mmult$ is matrix/vector multiplication and $\approx$ generalizes the case of scalar functions:
$$\lim_{||\dv|| \rightarrow 0} \frac{||f(v + \dv) - (f(v) + M \mmult \dv)||}{||\dv||} = 0$$
where $|| \ldots ||$ is the Euclidean norm.  We call $M_{ij}$, the $(i,j)$-th entry of $M$, the \emph{partial derivative of the $j$-th output of $f$ with respect to its $i$-th input at $v$}.

For example, for
$$f \begin{bmatrix} x_1 \\ x_2 \\ x_3 \end{bmatrix} = \begin{bmatrix} x_1 + x_2 \\ x_1 \cdot x_3 \end{bmatrix}: \Real^3 \rightarrow \Real^2$$
the matrix $\begin{bmatrix} 1 & 1 & 0 \\ -2 & 0 & 4 \end{bmatrix}$ is the derivative of $f$ at $\begin{bmatrix} 4 \\ 0 \\ -2 \end{bmatrix}$.

The \emph{Jacobi derivative of $f$} is the function $\deriv{f}$ that maps a vector $v$ to the Jacobi derivative of $f$ at $v$.  For example, for $f$ as above we have $$\deriv{f}(x_1, x_2, x_3) = \begin{bmatrix} 1 & 1 & 0 \\ x_3 & 0 & x_1 \end{bmatrix}.$$

The \emph{partial derivative of the $j$-th output of $f$ with respect to its $i$-th input} is the function
$\partial{f}_{ij}(v) = \deriv{f}(v)_{ij}$. This is usually written $\pderiv{f_j}{x_i}$ or even $\pderiv{y_j}{x_i}$.\footnote{We avoid this notation since the choi.ce of variable names for inputs and outputs of a function has nothing to do with the notion of derivative.}


The Leibniz derivative is the special case of a Jacobi derivative for $m = n = 1$.

\subsection{Fr\'echet derivative}

Jacobi derivatives are restricted to functions of the form $f: \prod{\seg{n}}{\Real} \rightarrow \prod{\seg{m}}{\Real}$, that is, finite-dimensional Euclidean spaces over the real numbers.
Sometimes it is convenient or even necessary to write functions where inputs are not tuples of scalars, but elements of possibly high-dimensional (or even infinite-dimensional) vector spaces.

\begin{example}
  \label{nn-layer}
  \memo{This example is now redundant and should probably be removed.}
A layer of a neural network is parameterized by a weight matrix $W \in \prod{\seg{m} \times \seg{n}}{\Real}$ and bias vector $b \in \prod{\seg{m}}{\Real}$ and takes a data vector $v \in \prod{\seg{n}}{\Real}$ as input, where $|\seg{m}|, |\seg{n}| \gg 0$ may be in the millions or billions.  It can be defined in a single line by
$$g(W,b,v) = \map{h} \, (W \mmult v + b)$$ where $\map{h}$ applies $h$ to each element of a vector, $h : \Func{\Real}{\Real}$ is an activation function such as $\tanh$, $\mmult$ is matrix/vector multiplication and $+$ is vector addition.  It can be straightforwardly evaluated using data-parallel implementations of these operations.
Trying to write such a function using only scalar variables would be a bad idea for multiple reasons.
\end{example}

\begin{definition}[Fr\'echet derivative]
For a function $f: V \rightarrow W$ on Banach spaces $V, W$, the \emph{linear function} $A \in \Hom{V}{W}$ is the \emph{Fr\'{e}chet derivative of $f$ at $v$} if it satisfies
$$f(v + \dv) \approx f(v) + A(\dv),$$ that is
\begin{equation*}
\lim_{||\dv||_V \rightarrow 0} \frac{||f(v + \dv) - (f(v) + A(\dv))||_W}{||\dv||_V} = 0 
\end{equation*}
where $|| \ldots ||_U$ is the norm that comes with the Banach space $U$.

The \emph{Fr\'{e}chet derivative of $f : \Func{V}{W}$} is the partial function $\deriv{f}: \Func{V}{(\Hom{V}{W})}$ that maps a vector $v \in V$ to the Fr\'{e}chet derivative of $f$ at $v$.
\end{definition}

\begin{example}
\label{nn-layer-derivative}
Consider the function
$$g_{W,b}(v) = \map{h} \, (W \mmult v + b).$$
Its Fr\'{e}chet derivative is
$$\deriv{g_{W,b}}(v) = \Delta (\map{\deriv{h}} \, (W \mmult v + b)) \lcomp (W \mmult)$$
where $\Delta (f_1, \ldots f_m) (x_1, \ldots, x_m) = (f_1(x_1), \ldots, f_m(x_m))$ is zip-apply and $\lcomp$ is linear function composition.
For $h = \tanh$ we have
$$\deriv{\tanh}(x) = ((1 - \tanh^2(x)) \cdot).$$
Note that $\deriv{\tanh}(x) : \Hom{\Real}{\Real}$, which explains the use of the section notation
$((1 - \tanh^2(x)) \cdot)$.  Since $\Hom{\Real}{\Real}$ is isomorphic to $\Real$, the Leibniz derivative is $\deriv{\tanh}(x) = 1 - \tanh^2(x)$ via implicit application of the isomorphism to its Fr\'echet derivative.\footnote{The pleasant compositional nature and applicability of Fr\'echet derivatives arrives from \emph{not} performing this isomorphism such that the chain rule is always functional composition of linear functions, no matter which vector space the arguments and results of functions belong to.}  In particular we can rewrite $\deriv{g_{W,b}}$ as
\begin{eqnarray*}
\deriv{g_{W,b}}(v) & = & \Delta (\map{\deriv{h}} \, (W \mmult v + b)) \lcomp (W \mmult) \\
    & = & \Delta (\map{(\lambda x. (1 - \tanh^2(x))} \cdot) (W \mmult v + b))\lcomp (W \mmult) \\
    & = & \zipWith (\cdot) (\map (\lambda x. (1 - \tanh^2(x))) (W \mmult v + b)) \lcomp (W \mmult)
\end{eqnarray*}
Note that the right-hand side is built by composing linear functions into a linear function: $\zipWith(\cdot)$ is bilinear and thus $\zipWith(\cdot) (\map (\lambda x. (1 - \tanh^2(x))))$ is linear; likewise, $\mmult$ is bilinear and thus the section $(W \mmult)$ defined by $(W \mmult)(v) = W \mmult v$ is linear; and functional composition of linear functions by $\lcomp$ preserves linearity.

We can $\eta$-expand the combinatory expression on the right-hand side into a more familiar looking term representation:
$$\deriv{g_{W,b}}(v)(\dv) = \zipWith (\cdot) \,\, (\map (\lambda x. (1 - \tanh^2(x))) (W \mmult v + b)) \,\, (W \mmult \dv)$$
This holds for any dimensions of $W$ and can be computed entirely symbolically, based on general differentiation rules for composition (chain rule), constant, linear and bilinear function, general second-order operators such as $\map$ and a dictionary of derivatives for primitive functions such as $\tanh$.
\end{example}

The Fr\'echet derivative generalizes the Jacobi derivative. If $M$ is the Jacobi derivative of $f$ at $x$ then $A = (M \mmult)$ is its
Fr\'echet derivative.



%

\subsection{Gateaux derivative}

\begin{definition}[Gateaux differential, Gateaux derivative]
For a function $f: V \rightarrow W$ on Banach spaces $V, W$ over field $K$ (either $\Real$ or $\Complex$), $v$ an interior point of $V$, $\dv \in V$ an
\emph{input differential}, the \emph{output differential} $\dy \in W$ is the \emph{Gateaux differential of $f$ at $v$ in the direction $\dv$} if 
$$\dy = \lim_{t \rightarrow 0} \frac{||f(v + t \cdot \dv) - f(v)||_W}{||t||_K}$$
where $|| \ldots ||_W$ is the norm that comes with the Banach space $W$ and the underlying field $K$, respectively.

The \emph{Gateaux derivative of $f : \Func{V}{W}$} is the partial function $\deriv{f}: \Func{V \times V}{W}$ that maps $v, \dv \in V$ to the Gateaux differential $\deriv{f}(v, \dv)$.

Function $f$ is \emph{Gateaux differentiable at $v$} if its Gateaux differential exists at $v$ for all directions $\dv$.
\end{definition}
Gateaux derivatives generalize directional derivatives in multivariate analysis analogous to Fr\'echet derivatives generalizing total derivatives. 
They are more general than Fr\'echet derivatives in the sense that a function may be Gateaux differentiable at $v$ without also being Fr\'echet differentiable, but if the Fr\'echet derivative exists then it determines the Gateaux derivative:
$$\deriv{f}_{\mathit{Gateaux}}(v, \dv) = \deriv{f}_{\mathit{Fr\acute{e}chet}}(v)(\dv).$$

There are more general notions of derivatives. Some distinguish the spaces of vectors and differentials, some apply to continuous functions that that are not conventionally differentiable everywhere such as the absolute-value function $f(x) = |x|$.  For the purposes of this paper, Fr\'echet and Gateaux derivatives are sufficient.

\subsection{Gateaux versus Fr\'echet derivatives for automatic differentiation}

Gateaux derivatives are conceptually the basis of \emph{forward-mode automatic differentiation}, since they give rise to interpreting a term (or program) representation of a function as operating on \emph{dual numbers/dual tensors} $(v, \dv)$:
$$f^{[\mathit{fad}]}_{\mathit{Gateaux}}(v, \dv) = (f(v), \deriv{f}(v, \dv)),$$
which preserves functional composition
$$(g \comp f)^{[\mathit{fad}]}_{\mathit{Gateaux}} = g^{[\mathit{fad}]}_{\mathit{Gateaux}} \comp f^{[\mathit{fad}]}_{\mathit{Gateaux}}$$
and is thus easy to implement by replacing the ordinary implementation of numbers and tensors by dual numbers and tensors, respectively.
This camouflages, however, that the first component, called the \emph{primal value}, of the output only depends on the primal value of the input and 
that the second component, called the \emph{tangent value}, always depends linearly on the input tangent value. 
These universal properties can be partially recovered by partially evaluating $f^{[\mathit{fad}]}$ with static $v$ and dynamic $\dv$, which, by definition of $f^{[\mathit{fad}]}$, always succeeds with statically computing the primal value and leaving a partially evaluated program representing the derivative behind, but rendered in an expressive programming language that does not inherently capture that this is always a \emph{linear} function.  For example, in elemental and tensor-based automatic differentiation, the partially evaluated output will result in a data structure corresponding to a \emph{computation graph} \cite{bauer1974computational,linnainmaaTaylorExpansionAccumulated1976,linnainmaa1970}, also called tape\footnote{It is sometimes referred to as a Wengert tape, which we find surprising since Wengert describes only forward-mode AD in his 2-page article \cite{wengert1964simple}.} or trace.  Reversing its edges amounts to forming the adjoint of the linear function represented by the computation graph.   

First employing Gateaux derivatives underlying the dual number/tensor interpretation of a program just to recover Fr\'echet derivatives by partial evaluation seems like an unnecessary detour. Following Henglein \cite{henglein2016b} and Elliott \cite{elliott2018simple}, we argue that Fr\'echet derivatives are better suited for both symbolic and automatic differentiation 
since they capture and reify that the output value of a function only depends on its input value and the output differential depends 
on both input value and input differential, but always linearly on the input differential:
$$f^{[\mathit{fad}]}_{\mathit{Fr\acute{e}chet}}(v) = (f(v), \deriv{f}_{\mathit{Fr\acute{e}chet}}(v)) \in W \times (\Hom{V}{W}).$$
Since the derivative is always a linear function, it can be represented in a combinatory \emph{domain-specific language (DSL)} that is closed under linear functions and thus syntactically guarantees linearity of all constructed functions. 
This facilitates not only universal applicability of properties of linear functions, such as 
$f(x + y) = f(x) + f(y)$ for any $f$, but also symbolically computing adjoints, which are only defined for linear functions.  Linear functions generated during differentiation or adjoint differentiation (generating the adjoint of the derivative during differentiation) can be represented as ordinary functions ($\lambda$-abstractions) \cite{elliott2018simple}, of course, but this eliminates the possibility of subsequent optimization using linear and tensor algebra.

Executing adjoint derivatives, possibly after algebraic optimization, as ordinary functions (programs) provides computation of ``cheap'' gradients \cite{griewank2012invented} for scalar functions.  We believe that the functional-analysis based approach including tensor products in this paper
provides an implementation and data structure framework for not only cheap parallel computation of gradients for scalar functions, but also for cheap adjoint derivatives for non-scalar functions, where tensor decomposition (formal tensor products), tensor contraction and relational reduction have important roles to play.

\section{Applications}
\label{applications}

We consider a series of examples of increasing complexity to illustrate how the functions are
represented in point-free notation and how their derivative is computed symbolically.

\begin{example}
\label{lnsin}
Let $h(x) = \ln (\sin x)$, that is $h = \ln \comp \sin$ in point-free notation.  Thus
$$\deriv{h}(x) = \deriv{\ln} y \lcomp \deriv{\sin}(x)$$ where $y = \sin x$ by Rule~\ref{chain-rule}, the chain rule.  Since
$\deriv{\ln} y = (\cdot \frac{1}{y})$ and $\deriv{\sin} x = (\cdot \, \cos x)$ we get
\begin{eqnarray*}
\deriv{h}(x) & = & \deriv{(\ln \comp \sin)}(x) \\
             & = & \deriv{\ln}(\sin x) \lcomp \deriv{\sin}(x) \\
             & = & (\frac{1}{\sin x} \,\cdot) \lcomp (\cos x \, \cdot) \\
             & = & (\frac{1}{\sin x} \cdot \cos x \,\cdot) \\
             & = & (\frac{\cos x}{\sin x} \cdot)
\end{eqnarray*}
\end{example}

\begin{example}
\label{stdex}
Consider $y = f(x_1,x_2) = \ln(x_1) + x_1 \cdot x_2  - sin(x_2)$ \cite[p.9]{adsurvey}.
It can be written in point-free form as
\begin{eqnarray*}
f & = & \ln \comp \pi_1 + \pi_1 \lift{\cdot} \pi_2 - \sin \comp \pi_2 
\end{eqnarray*}
where $\lift{\cdot}$ is $\cdot$ lifted to functions: $(f \lift{\cdot} g)(x) = f(x) \cdot g(x)$.
Employing our rules of differentation we obtain
\begin{eqnarray*}
\deriv{f}(x_1, x_2)
& = & \deriv{(\ln \comp \pi_1 + \pi_1 \lift{\cdot} \pi_2 - \sin \comp \pi_2)}(x_1, x_2) \\
& = & \deriv{(\ln \comp \pi_1)}(x_1, x_2) + \deriv{(\pi_1 \lift{\cdot} \pi_2)}(x_1, x_2) - \deriv{(\sin \comp \pi_2)}(x_1, x_2) \\
& = & \deriv{\ln}(\pi_1(x_1,x_2)) \lcomp \deriv{\pi_1}(x_1,x_2) + \\
&   & (\pi_1(x_1,x_2)) \,\cdot) \lcomp \deriv{\pi_2}(x_1,x_2)) + (\cdot \, \pi_2(x_1,x_2)) \lcomp \deriv{\pi_1}(x_1,x_2) - \\
&   & \deriv{\sin}(\pi_2(x_1,x_2)) \lcomp \deriv{\pi_2}(x_1,x_2) \\
& = & \deriv{\ln}(x_1) \lcomp \pi_1 + \\
&   & (x_1 \,\cdot) \lcomp \pi_2 + (\cdot \, x_2) \lcomp \pi_1 - \\
&   & \deriv{\sin}(x_2) \lcomp \pi_2 \\
& = & (\frac{1}{x_1}\cdot) \lcomp \pi_1 + \\
&   & (x_1 \,\cdot) \lcomp \pi_2 + (\cdot \, x_2) \lcomp \pi_1 - \\
&   & (\cos x_2 \,\cdot) \lcomp \pi_2
\end{eqnarray*}
The last line is easily transformed into a familiar looking expressing by recognizing that $\partial x_1 = (\lcomp \pi_1)$ and $\partial x_2 = (\lcomp \pi_2)$:
\begin{eqnarray*}
\deriv{f}(x_1, x_2)(\partial x_1, \partial x_2)
& = & (\frac{1}{x_1} \cdot) (\partial x_1) + (x_1 \,\cdot) (\partial x_2) + (\cdot \, x_2) (\partial x_1) - (\cos x_2 \,\cdot) (\partial x_2) \\
& = & \frac{1}{x_1} \cdot \partial x_1 + x_1 \cdot \partial x_2 + \partial x_1 \cdot x_2 - \cos x_2 \cdot \partial x_2
\end{eqnarray*}
In Leibniz notation, we can get the partial derivative $\frac{\deriv{f}(x_1,x_2)}{\partial x_1}$ by setting $\partial x_2 = 0$ and formally dividing the righthand side by $\partial x_1$. Aanalogously we can compute $\frac{\deriv{f}(x_1,x_2)}{\partial x_2}$.  Finally, writing $\partial f$ instead of $\deriv{f}$ we get the familiar looking:
\begin{eqnarray*}
\frac{\partial f (x_1, x_2)}{\partial x_1} & = & \frac{1}{x_1} + x_2 \\
\frac{\partial f (x_1, x_2)}{\partial x_2} & = & x_1 - \cos x_2
\end{eqnarray*}
The derivative for the unreadable version of $f$ is as follows:
\begin{eqnarray*}
\deriv{f}(x_1, x_2)
& = &  \deriv{((+) \comp ( (+) \times ((-) \comp \sin) ) \comp \langle \langle \ln \comp \, \fst, (\cdot) \rangle, \snd \rangle)}(x_1, x_2) \\
& = & (\blank \deriv{+} \blank) \comp ((\blank \deriv{+} \blank) \times (\deriv{-} \blank \comp \deriv{\sin} x_2) )
\comp \langle \langle \deriv{\ln} x_1 \comp \deriv{\fst} \blank, x_1 \deriv{\cdot} x_2 \rangle, \deriv{\snd} \blank \rangle \\
& = & (+) \comp ( (+) \times ((-) \comp (\cos x_2 \cdot) ) ) \comp \langle \langle (\frac{1}{x_1} \cdot) \comp \fst,
(+) \comp ( (x_2 \cdot) \times (x_1 \cdot) ) \rangle, \snd \rangle
\end{eqnarray*}
\end{example}

Since the right-hand side in this example is constructed from linear functions and
combinators that preserve linearity we can see that it is linear, as it should be.  (The derivative at any point is linear by definition.)
More specifically, the code for the derivative is \emph{parametric} in $x_1, x_2$: it is the same for each point $(x_1, x_2) \in \Real \times \Real$.  It is not a definitional requirement that the derivative at one point
have the same expression as the derivative at another point, but when it does it is useful in practice:
If we need to compute the derivative at many different points and the derivative is described by the same program at each point, it makes sense to optimize that program prior to executing it.  What makes this extra intriguing is that derivatives, always denoting linear functions, are essentially first-order data with strong algebraic properties admitting powerful optimizations.

\newcommand{\rot}{\mathbf{r}}
\newcommand{\point}{\mathbf{X}}
\newcommand{\raddist}{\mathbf{\kappa}}
\newcommand{\twopoint}{\mathbf{x}}
\newcommand{\camcent}{\mathbf{C}}
\newcommand{\prinpoint}{\mathbf{x_0}}
\newcommand{\obspoint}{\mathbf{m}}
\newcommand{\camparams}{\mathbf{P}}

\end{document}